\documentclass[twocolumn]{aastex631}

\begin{document}

\title{A Composite Broad-Line Region in SDSS J1609+4902: a Double-Peaked Disk component and a Gaussian Component}

\author[0000-0002-2581-8154]{Jiancheng Wu}
\affiliation{Department of Astronomy, School of Physics, Huazhong University of Science and Technology, Luoyu Road 1037, Wuhan, China}
\author[0000-0003-4773-4987]{Qingwen Wu$^*$}
\affiliation{Department of Astronomy, School of Physics, Huazhong University of Science and Technology, Luoyu Road 1037, Wuhan, China}
\author{Chen Hu}
\affiliation{Key Laboratory for Particle Astrophysics, Institute of High Energy Physics, Chinese Academy of Sciences, 19B Yuquan Road, Beijing 100049, People’s Republic of China}
\author[0000-0001-8879-368X]{Bing Lyu}
\affiliation{Kavli Institute for Astronomy and Astrophysics, Peking University, Beijing 100871, Peoples Republic of China}
\author{Hua-Rui Bai}
\affiliation{Key Laboratory for Particle Astrophysics, Institute of High Energy Physics, Chinese Academy of Sciences, 19B Yuquan Road, Beijing 100049, People’s Republic of China}
\author{Yi-Xin Fu}
\affiliation{Key Laboratory for Particle Astrophysics, Institute of High Energy Physics, Chinese Academy of Sciences, 19B Yuquan Road, Beijing 100049, People’s Republic of China}
\author{Yu Zhao}
\affiliation{Key Laboratory for Particle Astrophysics, Institute of High Energy Physics, Chinese Academy of Sciences, 19B Yuquan Road, Beijing 100049, People’s Republic of China}
\author[0000-0001-9449-9268]{Jian-Min Wang}
\affiliation{Key Laboratory for Particle Astrophysics, Institute of High Energy Physics, Chinese Academy of Sciences, 19B Yuquan Road, Beijing 100049, People’s Republic of China}
\author[0000-0002-2355-3498]{Xinwu Cao}
\affiliation{Institute for Astronomy, School of Physics, Zhejiang University, 866 Yuhangtang Road, Hangzhou 310058, People’s Republic of China}

\begin{abstract}
The profiles of broad emission lines in active galactic nuclei (AGNs) provide critical insights into the geometry and kinematics of the broad-line region (BLR), which in turn influence the uncertainties in estimating the masses of central supermassive black holes. In this study, we report the discovery of a low-luminosity AGN, SDSS J1609+4902, in which the H$\alpha$ line exhibits two distinct BLR components: a Gaussian component and an extremely broad double-peaked component. Follow-up observations conducted at the Lijiang Observatory in 2025 reveal that the line profile remains roughly unchanged, suggesting that this BLR structure may remain stable over a timescale of $\sim$10 years. We find that the size of the central Gaussian (Full Width at Half Maximum, FWHM$\sim 3000\,{\rm km\, s^{-1}}$) component is consistent with the classical reverberation mapping correlation. In contrast, the asymmetric double-peaked wing (FWHM$\sim 23,000\,{\rm km\, s^{-1}}$) likely originates from a disk-like BLR with an inner radius of approximately 70 gravitational radii. These results provide new constraints on the structure and dynamics of BLRs in AGNs and highlight the potential for long-term stability in such systems.
\end{abstract}


\section{Introduction} \label{sec:intro}
Active galactic nuclei (AGNs) are a class of luminous objects, which are powered by an accretion disk surrounding a central supermassive black hole \citep[SMBHs, e.g.,][]{Peterson1997}. In the unified model \citep{Antonucci1993}, AGNs can be divided into two types based on the appearance of broad emission lines in their optical spectrum (Type I), or not (Type II).  In the Type I AGNs, the broad lines can roughly be fitted with a Gaussian profile with full width at half maximum (FWHM) $\gtrsim1000\,{\rm km\, s^{-1}}$ \citep[e.g.,][]{Shen2011}. In type II AGNs, the broad lines are obscured by the putative geometrically thick torus \citep[e.g.,][]{Netzer2015}. It should be noted that some true type II AGNs may also exist which have only narrow lines without strong absorption. The formation of the broad line region (BLR) is still unclear, and these true type II AGNs normally have low luminosities, where the BLR may be absent\citep[e.g.,][]{Moran2000}.

The broad lines are widely believed to reprocess the AGN continuum. Using the reverberation-mapping (RM) technique, the BLR size can be estimated by the time delay in the responses of the different broad lines to the variations in the continuum \citep{Peterson1993}, which suggests that the BLR should be stratified. Furthermore, the line profiles are not always symmetric or single-peaked. Some AGNs that show prominent broad wings in their profile \citep[e.g.,][]{Sulentic2002, Popvic2004, Hu2012, Nagoshi2024}. Previous studies suggested that there also exists an intermediate-line region by analyzing the Balmer line profiles \citep[e.g.,][]{Sulentic2000}. \cite{Ferland1990} show that broad wings remain unchanged while the core luminosity increases when Mrk 590 suffered dramatic changes in continuum and line luminosities, which implies the complex BLR structure \citep[see also,][]{Hu2008, Hu2020, Feng2025}. 

\begin{figure*}[]
\centering
\includegraphics[scale=0.75]{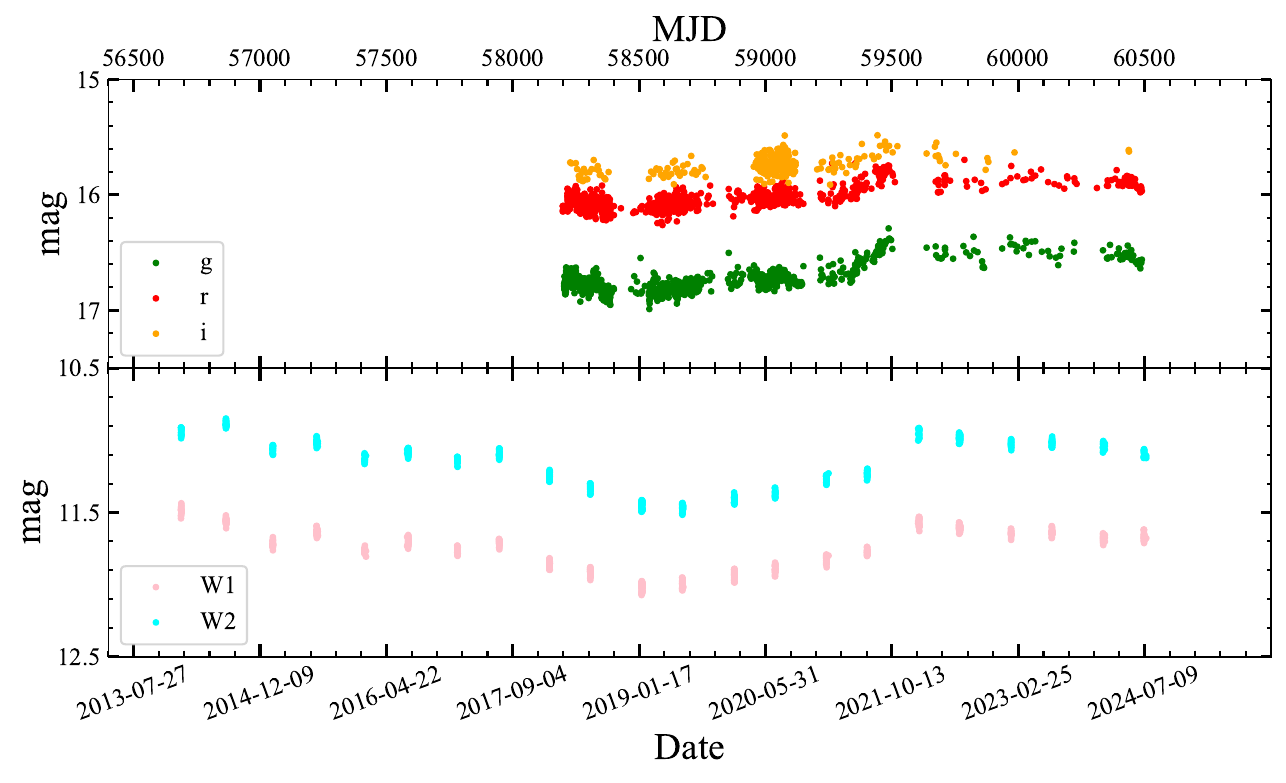}
\caption{The light curve for SDSS J1609 in optical and infrared wavebands, where the upper and lower panels show the ZTF $g$, $r$, $i$ bands and WISE W1, W2 wavebands respectively.
\label{fig1}}
\end{figure*}

Broad double-peaked Balmer emission lines are found in a fraction of AGNs, which are useful diagnostic for the structure and dynamics of the BLR \citep[e.g.,][]{Chen1989, Chen1989_2, Eracleous1994, Eracleous2003, Strateva2003}. These double-peaked lines are possibly associated with a flattened disk-like BLR structure \citep[e.g.,][]{Chen1989, Eracleous1995, Storchi2003}. The most simple one is a circular disk model, which predicts the double-peak profile with a stronger blue peak than the red one \citep[e.g,][]{Gilbert1999, Storchi-Bergmann2003, Schimoia2012}. It should be noted that the red peak is occasionally stronger than the blue one in some sources \cite[e.g., NGC 5548,][]{Li2016, Goad2016, Lu2022, Lu2024}. Therefore, the non-axisymmetric models are further proposed, which include either the circular disk with spiral arms \citep[e.g.,][]{Lewis2010, Storchi2003, Storchi2017, Schimoia2017} or elliptical disks \citep[e.g.,][]{Eracleous1995, Storchi1997}. These non-axisymmetric models can lead to modulations in the flux ratio of the red to blue peak in dynamical timescale \citep[e.g., NGC 5548][]{Li2016}. Apart from the disk-like BLR model, \cite{Zheng1990} also proposed that the double-peaked profile could be caused by the bi-conical outflow \citep[see also][for other anisotropy emission models]{Wanders1995, Goad1996}. In the time-domain era, more and more AGNs show strong variations, which provide a chance to further distinguish different models.

In this paper, we find a low-luminosity AGN ($z=0.04$) with two evident BLRs when searching for double-peaked AGNs. In this object, the extremely broad double-peak line may come from the disk emitter and the outer one follows the normal RM correlation. A flat ${\rm \Lambda CDM}$ cosmological model is adopted with $H_0=70 \,{\rm km\,s^{-1}\,Mpc^{-1}}$, $\Omega_{\rm m}=0.3$ and $\Omega_{\rm \Lambda}=0.7$.

\section{Observation and data reduction} \label{sec:data}

\subsection{Photometry}
The Zwicky Transient Facility \citep[ZTF,][]{Bellm2019} using an extremely wide field of view camera scans the Northern sky every two days. SDSS J1609+4902 (hereafter SDSS J1609) has been monitored since 2018 by ZTF in the $g$, $r$ and $i$ wavebands. Wide-field Infrared Survey Explorer (WISE) launched in December 2009 performs an all-sky survey every 6 months for four mid-IR wavebands 3.4, 4.6, 12, and 22 ${\rm \mu m}$ (refer to W1, W2, W3, and W4). WISE started a new mission Near Earth Object WISE (NEOWISE) after the frozen hydrogen cooling telescope was depleted and only the W1 and W2 bands were working \citep{Mainzer2014}. We retrieve the WISE photometry data in W1 and W2 wavebands for SDSS J1609. The light curves for these photometry data are presented in Figure \ref{fig1}, where outliers beyond 3$\sigma$ have been removed due to the possible measurement errors or artifacts. To build the multi-waveband spectrum, we also include the photometry in FUV and NUV wavebands by the Galaxy Evolution Explorer \citep[GALEX,][]{Martin2005} as shown in Figure \ref{fig3}.

\begin{deluxetable*}{lcc}[ht]
\centering
\tabcolsep=0.4cm
\tablecaption{Fitting parameters for the broad H$\alpha$ line observed by SDSS and LJT. \label{table1}}
\tablehead{
\colhead{parameter} & \colhead{SDSS} & \colhead{LJT}
}

\startdata
$q$ & $2.8 \pm 0.2$ & $2.2\pm0.6$\\
$i [^{\circ}]$ & $30.1 \pm 1.0$ & -\\
$\sigma [{\rm km\, s^{-1}}]$ & $919.6 \pm 282.4$ & $2016.9\pm254.0$\\
$R_{\rm in} [R_{\rm g}]$ & $69.8 \pm 4.4$ & $73.0\pm5.8$\\ 
$R_{\rm out} [R_{\rm g}]$ & $523.7 \pm 60.5$ & $400.6\pm26.0$\\
Gaussian $\lambda_{\rm peak} [{\text \AA}]$ & $6568.5 \pm 0.3$ & $6576.2\pm3.5$\\
Gaussian FWHM $[{\rm km\, s^{-1}}]$ & $3299.9 \pm 35.5$ & $3858.2\pm137.3$\\
\enddata 
\tablecomments{The inclination is kept unchanged for the LJT fitting as that in the SDSS.
}
\end{deluxetable*}

\begin{figure}[ht]
\centering
\includegraphics[scale=0.6]{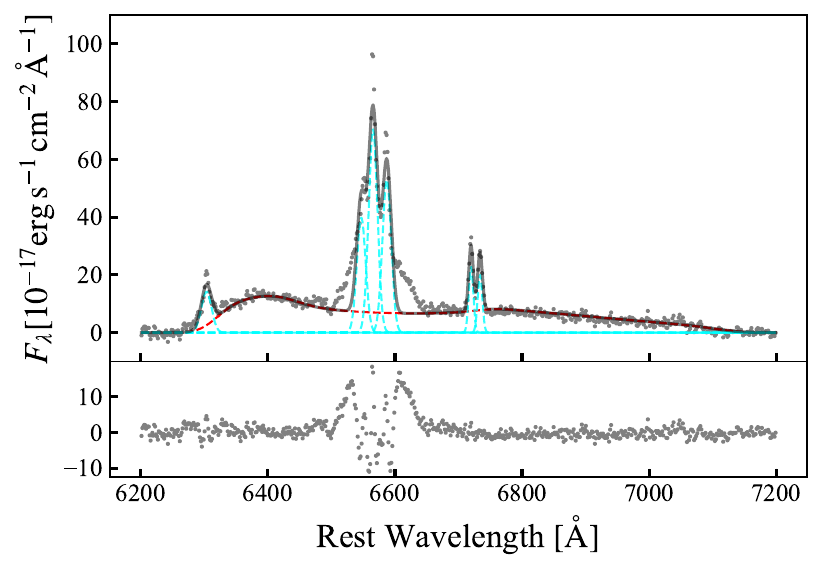}
\caption{The upper panel presents the fitting of the H$\alpha$ line with a broad double-peaked component (red dashed line) and six Gaussian components for narrow lines of H$\alpha$, [O {\footnotesize I}], [N {\footnotesize II}] doublet, and [S {\footnotesize II}] doublet (cyan dashed lines) for SDSS J1609. The sum of all components is represented by the gray line. The gray points represent the data after subtracting the continuum emission, where the fitting is not so good with the reduced chi-square $\chi^2_{\rm r}=11.3$. The lower panel presents the fitting residual.
\label{fig_resi}}
\end{figure}

\subsection{Spectroscopy}
Sloan Digital Sky Survey (SDSS) is a wide-field optical sky survey, and the wavelengths cover from 3800${\rm \AA}$ to 9200${\rm \AA}$ with spectral resolution $R\sim2000$. We retrieve the archived data from SDSS DR17 \citep{SDSS17}, where SDSS J1609 was observed on June 9, 2013 (MJD 56453). The SDSS pipeline classifies it as a galaxy with redshift $z=0.04$. We take a follow-up spectroscopic observation by the Lijiang telescope (LJT) on Feb 6, 2025, where the LJT is a 2.4 m telescope located at Lijiang Observatory \citep{Wang2019}. We use the Grism 8 (G8) which cover from 5100{\text \AA} to 9600{\text \AA} with dispersion 1.5{\text \AA} per pixel. We calibrate the LJT spectrum using the flux of the [S {\footnotesize II}] $\lambda\lambda 6718, 6732$ doublet.

\begin{figure*}[ht]
\centering
\includegraphics[scale=0.7]{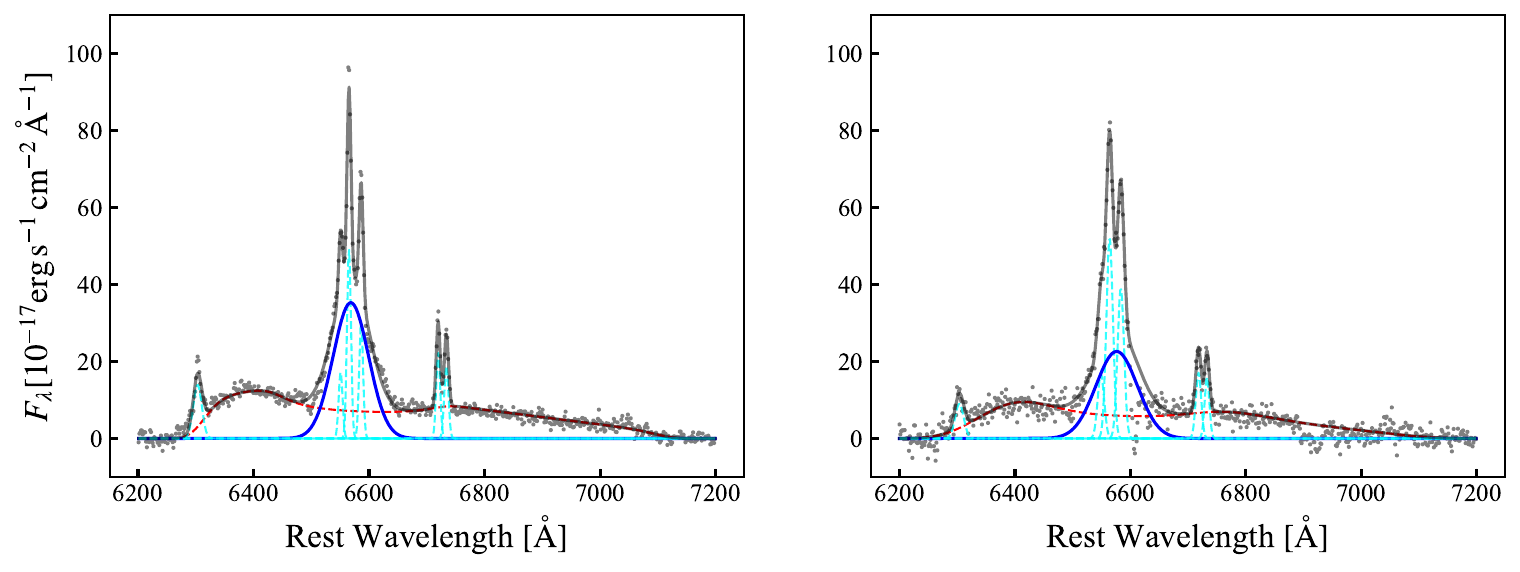}
\caption{The fittings for the broad H$\alpha$ line with a broad double-peaked component (red dashed line) and a Gaussian component (blue solid line). The gray points represent the data after subtracting the continuum emission. Narrow lines of H$\alpha$, [O {\footnotesize I}], [N {\footnotesize II}] doublet and [S {\footnotesize II}] doublet are fitted with Gaussian (cyan dashed lines). The sum of all components is represented by the gray line. In the left panel, the reduced chi-square $\chi^2_{\rm r}=1.4$ for the SDSS spectrum. In the right panel, the reduced chi-square $\chi^2_{\rm r}=5.3$ for the LJT spectrum with the atmospheric absorption masked.
\label{fig2}}
\end{figure*}

\begin{figure}[ht]
\centering
\includegraphics[scale=0.55]{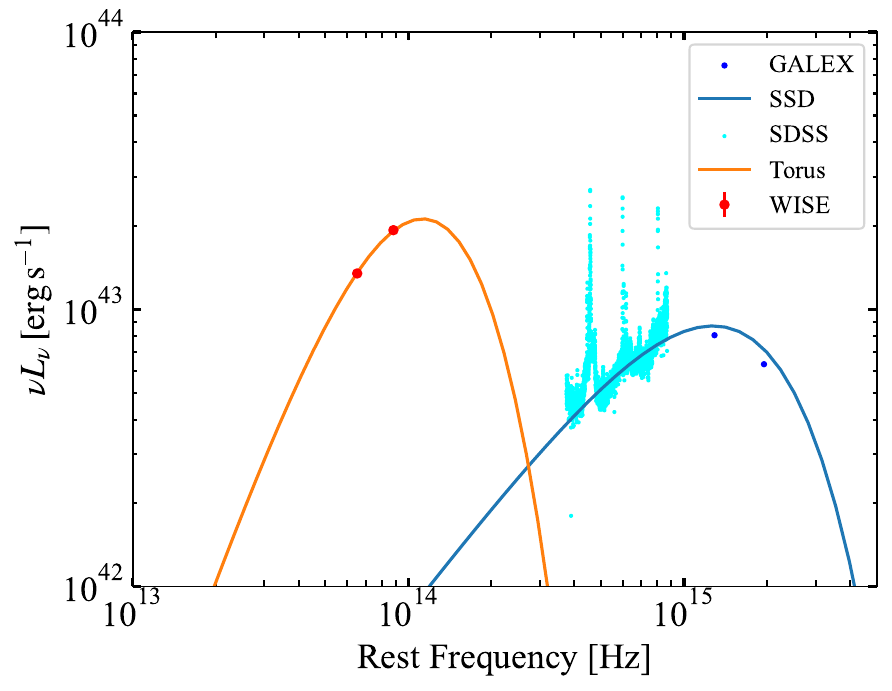}
\caption{The SED modeling. The temperature of the blackbody component is $1380\,{\rm K}$ for the near-infrared waveband. The optical-UV data is fitted with a truncated thin disk with a truncation radius of $40R_{\rm g}$ with $M_{\rm BH} = 3.3\times10^7M_{\odot}$ and $\dot{m}=0.017$.
\label{fig3}}
\end{figure}

\section{Analysis and Result}
\subsection{Infrared Time Delay}
To explore the time delay between the optical bands and mid-infrared bands, we use the NEOWISE photometry data and ZTF photometry data between MJD 58000 and 60000 to perform an interpolate cross-correlation function (ICCF) by utilizing the open source code PyCCF \citep{Sun2018, Shen2024}. Two methods of Flux Randomization and Random Subset Selection are employed with 2,000 Monte Carlo realizations to estimate the cross-correlation centroid distribution, with the lag range [-200, 200] in PyCCF. We find that the centroid time delay between the WISE W1 band and the ZTF g band is $52.0^{+6.0}_{-5.5}$ days. This time delay is roughly consistent with empirical correlation as found in the reverberation mapping \citep[e.g.,][]{Mandal2024}, see Figure \ref{fig4}.

\subsection{Spectrum Decomposition} \label{decomposition}
In low-luminosity AGNs, an important fraction of optical emission may be contributed by the host galaxy. In this work, spectrum decomposition is employed to get the AGN spectrum of SDSS J1609 by adopting the PyQSOFit \citep{Guo2018, Shen2019, Ren2024}. The stellar velocity dispersion could also be estimated by the absorption feature in the host galaxy spectrum given by PyQSOFit. We adopt `DZBIN1' and `PCA' for parameters `qso\_type' and `host\_type', and the numbers of each set of templates are 10 \citep{Yip2004}. We obtain the stellar velocity dispersion $\sigma_{*} = 137.7 {\rm km\, s^{-1}}$. The SMBH mass is  $3.3\times10^7 M_{\odot}$, $2.3\times10^7 M_{\odot}$ and $3.9\times10^7 M_{\odot}$ based on the empirical relations $M_{\rm BH}-\sigma_{*}$ in \cite{Woo2015}, \cite{Ferrarese2000} and \cite{Gebhardt2000} respectively. In this work, we simply adopt $3.3\times10^7 M_{\odot}$. 

\subsection{Emission Line Fitting}
The broad H$\beta$ line of SDSS J1609 is weak, and, therefore, we only focus on the H$\alpha$ line. In this work, we adopt the circular disk model in fitting of the broad H$\alpha$ profile \citep{Chen1989_2}, where the narrow components of H$\alpha$, [O {\footnotesize I}] $\lambda 6300$, [N {\footnotesize II}] $\lambda \lambda 6548, 6583$ and [S {\footnotesize II}] $\lambda \lambda 6719, 6734$ are also considered. The spiral arm or anisotropic hotspot is neglected due to the circular disk model can well reproduce the stronger blue peaks for SDSS J1609. We find that the residual is evident near the line center if only considering the broad disk component and narrow-line components (see Figure \ref{fig_resi}). Therefore, we further consider an additional broad Gaussian component near the central wavelength of the emission line. We present the fitting results of all the Gaussian component parameters in Table \ref{table1}. The fitting results of the BLR model are shown in Figure \ref{fig2}. It is noted that several studies have reported the presence of slightly broader and blue-shifted wing components in addition to the narrow cores for [O {\footnotesize III}] lines and some [S {\footnotesize II}] lines \citep[e.g.,][]{Whittle1985, Storchi1992, Wang2011, Singha2022}. If this is also the case for the narrow components of H$\alpha$ and [N {\footnotesize II}], the extra Gaussian component in our fitting should exhibit a blue shift. Furthermore, the wing components are also absent in the [S {\footnotesize II}] lines. Therefore, the extra central broad component is most possibly a broad-line component. For the very broad component, we simply estimate the FWHM of the asymmetric component by locating the half-maximum points of the peak, which is ${\rm FWHM}=22,901.1 \,{\rm km\, s^{-1}}$ \citep[see also,][]{Peterson2004}. The FWHM of the central normal broad Gaussian component is $3299.9\,{\rm km\,s^{-1}}$. Considering the inclination effect and assuming the BLR is virialized, we can estimate the size of the normal BLR $R\sim GM_{\rm BH}{\rm sin}^2 i/{\rm FWHM}^2$, which is 3.9 light days and roughly follow the typical RM correlation of $R_{\rm BLR}-L_{\rm bol}$ \citep[see Figure \ref{fig4}][]{Kaspi2005}. In the LJT spectrum, we find significant atmospheric absorption on the red side of the H$\alpha$ line, and we mask the data within the range of [6590, 6630]{\text \AA} and [6870, 7060]{\text \AA} in the spectral fittings.

\subsection{SED Modeling}
We model the broadband spectrum of SDSS J1609 using the classical models as widely adopted in AGNs. For infrared WISE W1 and W2 data, we simply use a blackbody to fit the putative torus emission, where we find the blackbody temperature is $\sim1380\, {\rm K}$ with a luminosity of $2.9\times10^{43} {\rm erg\,s^{-1}}$. For the optical to UV data, we find that the fitting is better with a truncated disk with a transition radius of $\sim40R_{\rm g}$ comparing the classical thin disk extended to several  $R_{\rm g}$ \citep{Shakura1973}, where the dimensionless mass accretion rate $\dot{m}=0.017$, see Figure \ref{fig3}.

\section{Conclusion and Discussion}
In this work, we report a low-luminosity AGN with two evident BLRs based on the H$\alpha$ line, where the inner BLR could be fitted with a disk BLR model while the outer BLR follows a single Gaussian profile. We find that the distance of these two BLRs may differ by a factor of 10. Our follow-up observation with LJT shows that the line profile shows little variation after 12 years.

\begin{figure}[ht]
\centering
\includegraphics[scale=0.57]{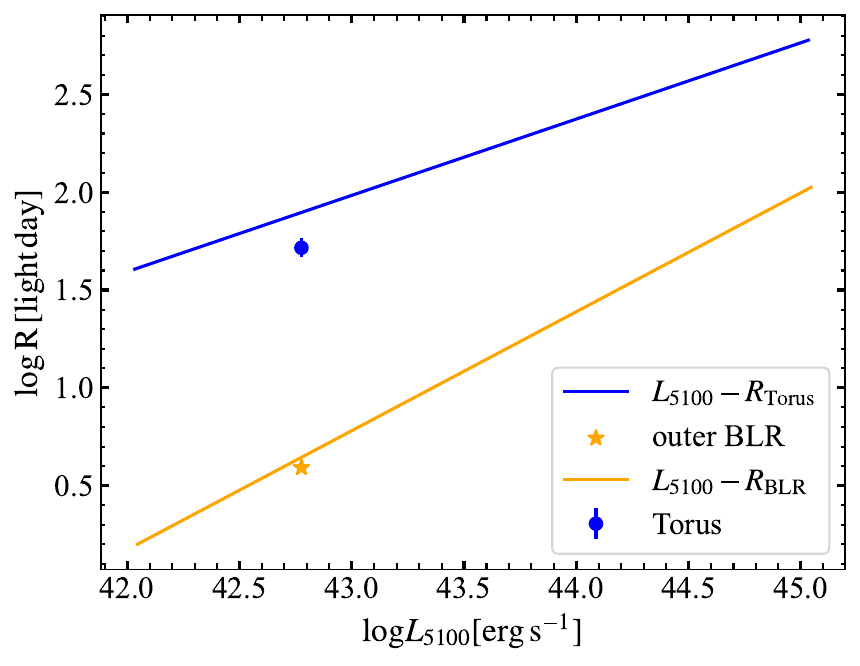}
\caption{The comparison of the size of torus (blue circle) and outer BLR (orange star) with the empirical correlations \citep{Kaspi2005, Mandal2024}, where the solid lines represent the observed empirical relation for torus (blue) and BLR (orange), respectively.
\label{fig4}}
\end{figure}

The existence of two separate BLRs has been proposed to reproduce the observed line profiles \citep{Netzer2010}. \cite{Wang2014} also suggested that the slim disk can separate the BLR into different regions in the super-Eddington regime, where the BLRs receive different ionizing emissions and arise different line profiles. For SDSS J1609, the accretion rate is low $\sim0.02$, which cannot lead to two BLRs as that in super-Eddington sources. The BLR clouds should be closely correlated with the accretion process if they are formed from or affected by the accretion disk. \cite{Kartje1999} proposed that the BLR structure will change with the decrease of the luminosity in disk-wind scenario. \cite{Elitzur2014} analyzed the evolution of broad lines along the evolution of accretion rate based on the disk-wind scenario \citep[][]{Elitzur2006, Elitzur2014}, where they found that a Gaussian component and double-peaked component will naturally appear in some intermediate state. This scenario can also explain the appearance or disappearance of emission lines in some changing-state quasars \citep[e.g.,][]{LaMassa2015}. \cite{Wu2024} simulated the BLR evolution in strong variable AGNs and found that two BLR components can co-exist at some certain accretion rate. For example, the former BLR still stays there and a new BLR may be formed during the AGN become brighten. \cite{Nagoshi2024} found two distinct regions in the BLR in a changing-state quasar SDSS J125809.31+351943.0, where the complex broad-line profile can be fitted with a single Gaussian component and a double-peak component. In their work, the inner BLR did not exhibit a response to variations in the continuum. For the outer BLR, the velocity-resolved reverberation mapping did not reveal any outflow signatures. They proposed that these two BLRs do not originate from outflows but from a dusty torus.

Two BLR components have been reported in some former works \citep[e.g.,][]{Strateva2003, Popvic2004}. For SDSS J1609, the inner BLR is extremely broad with ${\rm FWHM}\sim23,000 \,{\rm km\, s^{-1}}$ compared to the other reported sources, which may correspond to some special stage of BLR evolution. Such an extremely broad double-peaked profile has been reported in previous studies \citep[e.g.,][]{Wang2005}. The outer BLR component roughly follows the RM correlation, which suggests this component is a classical BLR. The extremely broad component follows a profile of disk emitter, which roughly extends to several tens gravitational radius. It should be noted that the line profile of H$\alpha$ is roughly unchanged in more than ten years. \cite{Ward2024} explored the variation of double-peaked profile and found that about half sources display significant changes over 10-20 years in strong variable AGNs. However, both the continuum and line profile in SDSS J1609 are roughly unchanged, which suggests that the line profile may be mainly regulated by the activity of central supermassive BHs in long timescale. Future more observations at shorter timescale can shed light on this issue.

In Figure \ref{fig4}, we compare the estimated size of the outer BLR and the torus with the empirical correlations derived from the RMs \citep{Kaspi2005, Mandal2024}. Both the location of the outer BLR and the torus are roughly consistent with their empirical correlations, respectively. This implies that the outer BLR is the normal one, and the inner disk BLR is an additional component. The size of the outer BLR is smaller than the inner radius of the torus. This is a little different from that in \cite{Nagoshi2024}, where they suggested that one BLR component may originate from the inner surface of the torus. The inner disk-like BLR ranges from 70 to 520$R_{\rm g}$ (or 0.13 to 0.98 light days), where the inner boundary of the disk-like BLR is more or less similar to the radius of the truncated disk (40$R_{\rm g}$), which may suggest that the disk-like BLR closely correlates with the properties of the outer thin disk. \cite{Nicastro2000} proposed that the BLR originates from accretion disk instabilities near the critical transition radius between gas and radiation pressure dominated regimes \citep{Nicastro2003}. For SDSS J1609, this mechanism would predict a BLR size of $\sim2500R_{\rm g}$ (or 4.7 light days), which is consistent with the outer BLR but much larger than the inner one. \cite{Czerny2011} attributed BLR formation to radiation-pressure-driven outflows from dust grains, which typically locate at distances of several thousand $R_{\rm g}$ from the central SMBH \citep{Naddaf2022}. It is still difficult to explain the extremely broad component in SDSS J1609 now. The high sensitivity and continuous spectral monitoring will help to better constrain the different components and further help to distinguish their origin.

\begin{acknowledgments}
The authors thank the referee for very constructive comments and suggestions. The work is supported by the National Natural Science Foundation of China (grants 12073023, 12233007, 12361131579, 12347103 and U1931203), the science research grants from the China Manned Space Project (No. CMS-CSST-2025-A07 and No. CMS-CSST-2021-A06) and the fundamental research fund for Chinese central universities (Zhejiang University). The authors acknowledge Beijng PARATERA Tech CO., Ltd. for providing HPC resources that have contributed to the results reported in this paper. 
\end{acknowledgments}

\vspace{5mm}

\software{PyCCF; Astropy\citep{Astropy1, Astropy2, Astropy3}}

\bibliography{sample631}{}
\bibliographystyle{aasjournal}



\end{document}